\documentstyle[12pt,amssymbols,epsf]{article}
\input math_macros.tex

\begin{document}
\begin{titlepage}
%\today                  \hfill
\begin{center}
\hfill    LBNL-40111 \\
%\hfill    UCB-PTH-97/xx \\
%\hfill hep-ph/97xxxxx\\
\vskip .5in

{\large \bf Removing flavor changing neutral interactions \\
      from leptoquark exchange}

\vskip 0.3in

Mahiko Suzuki

{\em Department of Physics and Lawrence Berkeley National Laboratory\\
University of California, Berkeley, California 94720}

\end{center}

\vskip .5in

\begin{abstract}
We look for a mechanism that removes without numerical fine tuning the strong 
constraint imposed by the flavor changing neutral current interaction of 
leptoquark exchange. If $n \times n$ degenerate leptoquarks couple
universally to $n$ generations of quarks and leptons in the weak basis,
not only can the  neutral current interaction be flavor diagonal, but also the 
charged current interaction takes exactly the same form as W exchange at 
low energies.  Atomic parity violation still imposes tight constraints.
Electroweak doublets with some left-right symmetry have the best
chance to be the lightest leptoquarks.   

\end{abstract}

\vskip .5in


\end{titlepage}

\renewcommand{\thepage}{\arabic{page}}

\setcounter{page}{2}

 Search of relatively light leptoquarks has been 
a subject of phenomenological interest. Such particles do not fit so well
to the popular grand unification \cite{GG} of the Standard Model, though
some of them behave like exotic colored Higgs bosons 
or R-parity violating squarks.
With SU(2)$\times$U(1) symmetry alone, the leptoquark exchange interaction 
generally allows chirality flip for both quarks and leptons. 
Furthermore, when fermion masses are diagonalized, leptoquark exchange can 
induce flavor changing neutral current (FCNC) interactions. 
Existing strong experimental bounds on chirality flip and 
FCNC interactions at low energies, particularly for the first and 
second generations of fermions, seem to rule out light leptoquarks that 
couple substantially to quarks and leptons: The typical constraint is 
$m_{LQ}/|g_{LQ}| \geq $ O(10 TeV) \cite{Shanker,Buch,Leurer}.  
If this is an unavoidable conclusion, future search of leptoquarks at HERA 
and LEP2 will be less motivated.

  It is easy to forbid chirality changing interactions by requiring 
that leptoquarks couple exclusively to one set of chiralities 
not to the other. On the other hand, it is less easy to 
eliminate the FCNC interaction, given the mass generation mechanism of 
the Standard Model. Is there any way to remove or suppress the FCNC 
interaction without making numerical fine tuning of the mass and 
coupling parameters ?  We look for such mechanism or symmetry scheme and
reexamine the bounds on leptoquark parameters in this 
paper.  Our purpose is not to propose a realistic or aesthetic extension of the
Standard Model, but rather to make a bottom-up phenomenological study as to
how far leptoquark masses can be lowered without contradicting existing
experiment.

   We shall attempt to remove the FCNC interactions, following the path of
the Glashow-Iliopoulos-Maiani (GIM) mechanism for Z and $\gamma$.
In the case of leptoquarks, we must introduce a certain universal coupling 
pattern as a hypothesis or a consequence of a global symmetry, while such 
universality is automatic for the gauge couplings of the GIM mechanism. 
Though our scheme may look less natural in this respect, 
it appears to be the only mechanism that removes the FCNC interaction without 
resort to numerical fine tuning of parameters.  Once our mechanism has been 
built in, we find the four-fermion interaction of leptoquark exchange 
after the Fierz rearrangement to be of the current-current form:
\begin{equation}
    L_{eff}  \sim \sum J_{\mu}^{(q)\dagger}J^{(\ell)\mu} + h.c.,
\end{equation}
where quarks and leptons enter the currents in a {\it generation independent} 
manner up to the Cabbibo-Kobayashi-Maskawa mixing matrix $V_{CKM}$ in the 
charged current. This interaction hides the dangerous interactions most 
effectively: For instance, it has no effect on lepton universality.
It affects only the overall scale of the charged current 
interaction and the neutral current processes off the Z mass peak. 
Neutrino scattering and atomic parity violation are among the most
sensitive of the latter.  One obvious way to avoid the neutral 
current constraints is to introduce left-right symmetry of one kind 
or another. 
 
     We study scalar leptoquarks here since they are more appealing than 
vector leptoquarks from the viewpoint of underlying renormalizability of theory.
Our mechanism can easily be extended to leptoquarks of higher spin.
      With SU(2)$\times$U(1) symmetry, the leptoquarks can come in a singlet 
($S, S', \tilde{S}$), a doublet ($R, R', \tilde{R}$) or a triplet 
(${\bf T}$) of the electroweak SU(2). They are all nonhermitian.
In the weak eigenstate basis, 
the SU(2)$\times$U(1) invariant couplings to quarks and leptons take the 
following form in an obvious notation:
\begin{eqnarray}
    L &=&   g\overline{U^c}_RE_RS
          + g'\overline{D^c}_RE_RS' 
          + (\tilde{g}/\sqrt{2})\overline{Q^c}_LL_L\tilde{S} \nonumber \\
             &+& h\overline{U}_RL_LR + h'\overline{D}_RL_LR'
       + \tilde{h}\overline{Q}_LE_R\tilde{R} 
       + (f/\sqrt{2})\overline{Q^c}_L\mbox{\boldmath$\tau$}L_L{\bf T} + h.c.,
                              \label{interaction}
\end{eqnarray} 
where the superscript $c$ for charge conjugation includes $i\tau_2$ for 
doublets and generation indices have been suppressed. Since each fermion 
carries one generation index, the couplings, $g, g', \tilde{g}, h, h', 
\tilde{h}$ and $f$, carry a pair of generation indices, one referring 
to the quark generation and the other to the lepton generation. 
We introduce just as many leptoquarks as the number of 
generation components of couplings.  That is, there are 
$n\times n$ distinct leptoquark fields for each species ({\it e.g.,} $S_A$ 
where $A=(a,b)$ with $a,b=1,2,\cdots n$) of leptoquarks plus their 
hermitian conjugates. With full indices, therefore, the couplings are label 
as $g_{cd}^A$ {\it etc}.  We shall see that this proliferation of 
leptoquarks is necessary for removing FCNC interactions.  We require 
also that $S$ and $\tilde{S}$ are distinct leptoquarks though they carry 
identical SU(2)$\times$U(1) quantum numbers. So are $R'$ and $\tilde{R}$. 
If they were identical, chirality changing interactions would
arise and disturb the flavor-diagonal decay 
$\pi\rightarrow e\nu_e$.

  We examine the FCNC interactions in two separate groups. 
The first group consists of $S, S', R, R'$ and $\tilde{R}$, while
$\tilde{S}$ and $T$ make the second group. The first group generates only
neutral current interactions in contrast to the second group which
leads to both neutral and charged current interactions.
Let us start with the first group. To be concrete, we work with the leptoquark
$S$. Instead of writing all generation indices 
explicitly, we express the $S$ interaction in a compact form by introducing
an $n\times n$ matrix $\cal G$ for product of the coupling $g$ and the field 
$S$:
\begin{equation}
           {\cal G}_{ab} = g_{ab}^A S_A,    
\end{equation}
where $A$ is summed over.
Since $\overline{U^c}_R$ and $E_R$ are {\it n}-component row and 
column vectors of generation, the $S$ coupling can be expressed simply as
      $ L = \overline{U^c}_R {\cal G} E_R. \label{S'1}$
When the low-energy limit of the $S$ propagator matrix
$ i\langle 0|T({\cal G}_{ab}(x){\cal G}_{cd}^{\dagger}(y))|0\rangle $  
is denoted by $\Delta_{ab,cd}$ in momentum space, the low-energy four-fermion 
interaction of $S$ exchange is written in the weak basis as
\begin{equation}
       L_{eff}= \sum_{a,b,c,d}
                 (\overline{U^c}_{Ra}E_{Rb})^{\dagger}\Delta_{ab,cd}
                         (\overline{U^c}_{Rc}E_{Rd}).
\end{equation}
When quarks and leptons are transformed from the weak basis (capital letters) 
to the mass basis (lower cases) by
the rotation $U_R = V_{uR}u_R$ and $E_R = V_{\ell R}e_R$, $L_{eff}$ turns into 
\begin{equation}
     L_{eff} =   \sum_{all} 
                (\overline{u^c}_{Ra'}V_{uR,a'a}^T
                V_{\ell R,bb'}e_{Rb'})^{\dagger}{\Delta_{ab,cd}}
                (\overline{u^c}_{Rc'}V_{uR,c'c}^T
                V_{\ell R,dd'}e_{Rd'}).   \label{Leff}
\end{equation}
We want this interaction to be diagonal in flavor, not by numerical
fine tuning of the coupling matrix $g$ and the $S$ masses to the rotation 
matrices $V_{uR}$ and $V_{\ell R}$, but by some mechanism that does not depend 
on specific details of the rotation matrices. The simple and only conceivable
solution is to exploit unitarity of $V_{uR}$ and $V_{\ell R}$ by choosing
mass and coupling of leptoquarks to be universal as specified 
in the following. First we require that the n$\times$n leptoquarks be 
all degenerate.  Next we choose their couplings such that in the weak basis, 
all transitions between a pair of the quark generation $a$ and the lepton 
generation $b$ are allowed only for $A=(a,b)$ with a universal strength 
independent of $a$ and $b$, {\it i.e.,} 
$g_{ab}^A=\delta_{A,ab}g$.  In other words, $S_A$ with $A=(a,b)$ 
mediates only between $U^c_{Ra}$ and $L_{Rb}$ with strength $g$. 
On this set of assumptions the low-energy $S$ propagator 
matrix takes the form:
\begin{equation}
            \Delta_{ab,cd} = \delta_{ac}\delta_{cd}
                       \biggl(\frac{g^2}{m^2_{S}}\biggr), \label{choice}    
\end{equation} 
where $g$ and $m_S$ are numbers.
$L_{eff}$ is now flavor diagonal in the mass basis: 
\begin{eqnarray}
       L_{eff} &=&\biggl(\frac{g^2}{2m^2_{S}}\biggr)
                (\overline{u^c}_R V_{uR}^T\gamma_{\mu}V_{uR}^{T\dagger}u^c_R)
                (\overline{e}_RV_{\ell R}^{\dagger}\gamma^{\mu}V_{\ell R}e_R), 
                          \nonumber \\
               &=&-\biggl(\frac{g^2}{2m^2_{S}}\biggr)
                  (\overline{u}_R\gamma_{\mu} u_R)
                  (\overline{e}_R\gamma^{\mu} e_R),
\end{eqnarray}
where summation over families is implicit for quarks and leptons separately.
Unlike the GIM mechanism, our mechanism must introduce the universal coupling
as a postulate. In addition, it requires that all generation components of $S$ 
be degenerate in mass. To make these requirements less {\it ad hoc}, 
we can introduce a family symmetry. The required pattern of coupling
universality and mass degeneracy both result if we introduce 
${\rm SU(n)}_q\times {\rm SU(n)}_{\ell}$ global family (generation) symmetry  
and assign the $S$ leptoquarks to its (n,n) representation. This symmetry
cannot be exact since it is explicitly broken by the Higgs interaction. 

  We can repeat the preceding argument for $S'$ by replacing the up-quarks 
by the down-quarks. Our mechanism works equally well for all other 
leptoquarks of the first group. Therefore $L_{eff}$ for exchange of 
$S, S', R, R'$ and $\tilde{R}$ is
\begin{eqnarray}
  L_{eff}^{(1)}  &=& -\biggl(\frac{g^2}{2m_{S}^2}\biggr)
     (\overline{u}_R\gamma_{\mu}u_R)(\overline{e}_R\gamma^{\mu}e_R) -
                \biggl(\frac{g'^2}{2m^2_{S'}}\biggr)
     (\overline{d}_R\gamma_{\mu}d_R)(\overline{e}_R\gamma^{\mu}e_R)
                      \nonumber \\ 
           &+& \biggl(\frac{h^2}{2m_{R}^2}\biggr)
     (\overline{u}_R\gamma_{\mu} u_R)(\overline{\ell}_L\gamma^{\mu} \ell_L) +
               \biggl(\frac{h'^2}{2m^2_{R'}}\biggr)
     (\overline{d}_R\gamma_{\mu} d_R)(\overline{\ell}_L\gamma^{\mu}\ell_L)  
            \nonumber \\
           &+&  \biggl(\frac{\tilde{h}^2}{2m^2_{\tilde{R}}}\biggr)
     (\overline{q}_L\gamma_{\mu} q_L)(\overline{e}_R\gamma^{\mu}e_R).
\end{eqnarray} 

  For the second group of leptoquarks, we notice that $\tilde{S}$ 
and $T_3(I_3=0)$ couple to two distinct charge states of quark-lepton as 
\begin{equation}
  L = (\tilde{g}/\sqrt{2})(\overline{D^c}_L N_L-\overline{U^c}_L E_L)\tilde{S}
   + (f/\sqrt{2})(\overline{D^c}_L N_L + \overline{U^c}_L E_L)T_3 + \cdots,    
\end{equation}
where suppressed are the couplings of $T(I_3=\pm 1)$. Let us make the 
postulate of Eq.(\ref{choice}) for $\tilde{S}$ and $T$ too. We bring the 
weak eigenstates to the mass/flavor eigenstates by the rotations;
\begin{equation}
    U_L = V_{uL}u_L, \;\;  D_L = V_{dL}d_L; \;\;
    E_L = V_{\ell L}e_L, \;\; N_L = V_{\ell L}\nu_L,
\end{equation}
where $\nu_L$ are in the electron, muon and tau-flavor basis, not necessarily
mass eigenstates if any of them has mass.  In the mass/flavor basis, the
effective four-fermion interaction takes the form
\begin{eqnarray}
   L_{eff}^{(2)} &=& -\biggl(\frac{\tilde{g}^2}{4m^2_{\tilde{S}}}\biggr)
              \biggl((\overline{u}_L\gamma_{\mu}\overline{u}_L)
                     (\overline{e}_L\gamma_{\mu}e_L)   
              +(\overline{d}_L\gamma_{\mu}d_L)
               (\overline{\nu}_L\gamma_{\mu}\nu_L) \nonumber \\
         &-& (\overline{d}_L\gamma_{\mu} V_{CKM}^{\dagger} u_L)
                  (\overline{\nu}_L\gamma^{\mu}e_L) -
              (\overline{u}_L\gamma_{\mu} V_{CKM} d_L)
                  (\overline{e}_L\gamma^{\mu}\nu_L)\biggr),  \label{Leff2}
\end{eqnarray}
where $V_{CKM}= V_{uL}^{\dagger}V_{dL}$ has been used.  The third and fourth 
terms of (\ref{Leff2}) are exactly of the same form as the charged current 
interaction of W exchange.  As far as the charged current part is 
concerned, this is the best we can do in hiding leptoquark interactions.
Its only consequence is in a shift of the effective Fermi decay constant 
$G_F$ for semileptonic processes. This would not be the case if $S$ 
and $\tilde{S}$ were identical
particles.  The $T$ exchange contribution is obtained similarly:
\begin{eqnarray}
  L_{eff}^{(3)}     &=& -\biggl(\frac{f^2}{4m^2_T}\biggr)
   [2(\overline{u}_L\gamma_{\mu}u_L)(\overline{\nu}_L\gamma_{\mu}\nu_L)
    +(\overline{u}_L\gamma_{\mu}u_L)(\overline{e}_L\gamma^{\mu}e_L)
                     \nonumber \\
    &+&(\overline{d}_L\gamma_{\mu}d_L)(\overline{\nu}_L\gamma^{\mu}\nu_L)
   +2(\overline{d}_L\gamma_{\mu}d_L)(\overline{e}_L\gamma^{\mu}e_L)
                     \nonumber\\ 
   &+&(\overline{d}_L\gamma_{\mu}V^{\dagger}_{CKM}u_L)
    (\overline{\nu}_L\gamma^{\mu}e_L) 
   +(\overline{u}_L\gamma_{\mu}V_{CKM}d_L)(\overline{e}_L\gamma^{\mu}\nu_L)].
                     \label{Leff3}
\end{eqnarray}
   Thus the total effective four-fermion interaction 
$L_{eff} = L^{(1)}_{eff} + L^{(2)}_{eff} + L^{(3)}_{eff}$ of leptoquark
exchange is not only chirality conserving but also flavor diagonal for the
neutral current part.

    We compare $L_{eff}$ with experiment.    Leptoquark 
exchange can alter the W and Z exchange amplitudes if leptoquarks 
are light enough and/or their couplings are strong enough. The effective 
charged current interaction of low-energy semileptonic processes is shifted 
in magnitude. Such a shift should be observable relative to
pure leptonic processes which are unaffected.  However, simple comparison of
$G_{\beta}$ for $\beta$ decay and $G_{\mu}$ for ${\mu}$ decay gives us no
information since the ratio $G_{\beta}/G_{\mu}$ is used to determine 
the CKM matrix element $V_{ud}$. It is futile to compare $G_{\beta}$ with 
the $W^{+}\rightarrow u\overline{d}$ decay rate since we do not 
have high enough precision for the latter. The best constraint of charged 
currents comes from unitarity of the CKM matrix. For neutral currents, 
leptoquark exchange alters low-energy semileptonic processes both in magnitude 
and in structure. Since the low-energy neutral weak current has been measured 
sufficiently accurately in a good agreement with the Standard Model 
prediction of Z exchange, it sets tight constraints on the 
leptoquark parameters. 

Let us start with the charged current. The CKM matrix elements $V_{ud}$, 
$V_{us}$ and $V_{ub}$ have been determined from $G_{\beta}/G_{\mu}$, 
$K_{e3}$ and the charmless semileptonic B decay, respectively.  
The values determined from these sources without a unitarity constraint 
add up very closely to unity \cite{PDG}:
\begin{equation}
     0.9919 < |V_{ud}|^2 + |V_{us}|^2 + |V_{ub}|^2 < 1.0008.  \label{CKM}
\end{equation}  
These bounds set the constraint:
\begin{equation}
        - 0.0081\times\frac{G_F}{\sqrt{2}}<
        - \frac{\tilde{g}^2}{8m^2_{\tilde{S}}}+ \frac{f^2}{8m^2_{T}} < 
         0.0008\times\frac{G_F}{\sqrt{2}}. \label{beta}
\end{equation} 
If either $\tilde{S}$ or $T$ exists, not both, then the bound is: 
\begin{equation}
      m_{\tilde{S}}/|\tilde{g}|>1.4 {\rm TeV},\; 
      m_{T}/|f| > 4.4 {\rm TeV}.   \label{STbound} 
\end{equation}
It is possible to eliminate the leptoquark effect in $G_F$ by cancellation 
between $\tilde{S}$ and $T$, {\it i.e.,} $|\tilde{g}|=|f|$. This is 
accomplished if we elevate the global electroweak SU(2) to global 
${\rm SU(2)}_{qL}\times{\rm SU(2)}_{\ell L}$
and assign ($\tilde{S}, {\bf T}$) to the (2,2) representation. Under this global
symmetry $S$ and $\tilde{S}$ are necessarily distinct particles.  So are $R'$ 
and $\tilde{R}$. If we wish the Higgs interaction to respect this symmetry, 
we would have to introduce at least two Higgs doublets, $H$=(2,1) 
and $H'$=(1,2), which generate mass for quarks and leptons separately. 

According to analysis of the neutral current data off the $Z$ peak,
a possible deviation of the $\nu u_L$ interaction from the Standard Model is
given by 
\begin{equation}
  - 0.0293  < \epsilon_L(u)|_{exp} - \epsilon_L(u)_{SM} < 0.0033 \label{uL}
\end{equation}
in terms of the parameter $\epsilon_L(u)$ \cite{PDG2}.
It leads us to the bound 
     $f^2/8m^2_T <  3.3\times 10^{-3} (G_F/\sqrt{2})$,  namely,
\begin{equation}
     m_{T}/|f| > 2.1 {\rm TeV}.  \label{mT}
\end{equation}
Similarly we obtain from $\epsilon_L(d)$ a bound for $\tilde{S}$:
\begin{equation}
     m_{\tilde{S}}/|\tilde{g}| > 3.1 {\rm TeV}.
\end{equation}
Though these tight bounds are free from uncertainties of accidental 
cancellation between terms, they are sensitive to the theoretical 
and experimental error estimates in the $\epsilon$ parameter values. 
We can go on to set a bound on $R$ from $\epsilon_R(u)$:
\begin{equation}
     m_{R}/|h| > 0.66 {\rm TeV}.
\end{equation}
There is not enough experimental accuracy in $\epsilon_R(d)$ to 
set a meaningful bound on $m_{R'}/|h'|$. $S$, $S'$ and $\tilde{R}$ have no
neutrino interaction. 

Parity violation in polarized eN scattering can also constrain 
the leptoquark couplings through the axial-vector 
current of electron.  Though the polarized $eN$ scattering has historical 
significance, advancement in theoretical and experimental precision of
atomic parity violation can now set a tighter constraint 
on the electron axial-vector current off the $Z$ mass peak \cite{Leurer}.  
The Cs experiment measures the weak charge $Q_W$:
\begin{equation}
    Q_W(Cs)= Q_W^{SM}(= -72.88 \pm 0.05 \pm 0.03)+\Delta Q_W,
\end{equation}        
where $\Delta Q_W$ is the possible leptoquark contribution:
\begin{eqnarray}
 \Delta Q_W &=& -\frac{2\sqrt{2}}{G_F}\Biggl((2Z+N)\biggl(\frac{g^2}{m^2_S}-
           \frac{\tilde{g}^2}{2m^2_{\tilde{S}}}+\frac{h^2}{m^2_R} -
           \frac{\tilde{h}^2}{m^2_{\tilde{R}}}-\frac{f^2}{2m^2_T}\biggr) + 
                    \nonumber \\
           &+& (Z+2N)\biggl(\frac{g'^2}{m^2_{S'}}+
              \frac{h'^2}{m^2_{R'}}-\frac{\tilde{h}^2}{m^2_{\tilde{R}}}-
              \frac{f^2}{m^2_T} \biggr)\Biggr),  \label{DeltaQ}
\end{eqnarray}  
with Z = 55 and N = 77.9.
The measured value \cite{Cs} is $Q_W(Cs)=-71.04\pm 1.58 \pm 0.88$, where 
the second error is theoretical uncertainties involved in atomic physics
calculation \cite{Johnson}. From $ -0.7 < \Delta Q_W < 4.4 $, we obtain the
following bounds:
\begin{equation}
       \frac{m_{LQ}}{|g_{LQ}|} > 2.9 {\rm TeV} \; (S, S', R, R'),
                       \label{Parity1} 
\end{equation}
if $\tilde{S}, \tilde{R}$ and $T$ are absent, and
\begin{equation}
       \frac{m_{LQ}}{|g_{LQ}|} > 0.8 {\rm TeV}\;  (\tilde{S}),\; %\nonumber\\   
                              1.7 {\rm TeV} \; (\tilde{R}),\; %\nonumber\\
                              1.4 {\rm TeV} \; (T),  \label{Parity2}
\end{equation}
if $S, S', R$ and $R'$ are absent. The constraint of atomic parity violation 
can be made weaker or even disappear if an appropriate symmetry 
is introduced for the mass and coupling of leptoquarks.
Since the four-fermion terms enter $L_{eff}$ in both signs, their sum 
cannot be brought into a pure vector-vector form.  However, we can turn 
it into the form $V_{\mu}V^{\mu}+A_{\mu}A^{\mu}$ by introducing right-handed 
neutrinos a left-right reflection symmetry: $Q_L \leftrightarrow Q_R$ 
and {\it simultaneously} $L_L\leftrightarrow L_R$. There
are two sets of coupling relations that satisfy this symmetry:  
\begin{eqnarray}
     g &=& -\tilde{g}/\sqrt{2}, \;\; m_S = m_{\tilde{S}};\;\;  %\nonumber \\ 
     g' = 0 = f; \nonumber \\
     h &=& h' = \tilde{h}, \;\; m_R = m_{R'} = m_{\tilde{R}}, \label{LR1}
\end{eqnarray}
and
\begin{eqnarray}
     g &=&  - g'/\sqrt{2} = - f/\sqrt{2},\;\; m_S = m_{S'}=m_T ;  \nonumber \\
     \tilde{g} &=& 0 ;\;\; %\nonumber \\
     h = h' = \tilde{h}, \;\; m_R = m_{R'} = m_{\tilde{R}}. \label{LR2}
\end{eqnarray}
In either case, parity violation is removed from the neutral current 
interaction, and the constraints of (\ref{Parity1}) and (\ref{Parity2}) 
become mute. However, $S$ and $S'$ are degenerate with $\tilde{S}$ or $T$, 
or else absent. Since $\tilde{S}$ and $T$ are 
constrained by the charge current, $R$, $R'$ and $\tilde{R}$ are left as 
the least constrained leptoquarks in mass and coupling.\footnote{
The R-parity violating squarks lead to a structure similar 
to the choice of $h'=\tilde{g}/\sqrt{2}$ and all other couplings 
equal to zero, apart from difference in the family multiplicity.
It is interesting to note that they escape from the atomic parity 
violation constraint by a large cancellation between the {\it u} and 
{\it d}-quark contributions in Eq.(\ref{DeltaQ}).} 
Direct production at HERA will be the most obvious test for 
light leptoquarks. Since the leptoquark exchange contribution 
to the $e^+e^-\rightarrow \overline{q}q$ does not fall 
off with center-of-mass energy, LEP2 will be competitive in leptoquark search. 

A remark is in order on the off-diagonal leptoquark masses
which can arise from the Higgs condensation. For instance, 
the global symmetry $SU(2)_{qL}\times SU(2)_{\ell L}$ forbids 
transition between $S$ and $\tilde{S}$, but the interaction 
$\lambda S^{\dagger}\tilde{S}H^{\dagger}H'$ generates the transition mass
$\lambda\langle H^{\dagger}\rangle\langle H'\rangle$. This 
transition is potentially dangerous since it leads to the chirality changing 
interaction $(\overline{Q}_LD_R)(\overline{E}_RL_L)$. The 
off-diagonal mass square must therefore be much smaller 
than $m^2_{LQ}$ in order to suppress chirality flip sufficiently.  
Need of this condition is quite general for 
low mass leptoquarks.  If we wish to lower the leptoquark mass down to
the electroweak scale, therefore, we must require either that the coupling 
$\lambda$ be very small or that $S$ and $\tilde{S}$ not coexist. 
The same is true for $R'$ and $\tilde{R}$.     

 To conclude, it is possible to remove flavor changing neutral current
interactions of leptoquark exchange, if multiple leptoquarks are introduced 
in a family symmetric way. Furthermore, if parity violation is removed 
from the neutral currents, leptoquarks can be made light and substantial 
in coupling. The lightest possible of such scalar leptoquarks is 
electroweak doublets which are pairwise degenerate under left-right reflection.
Each doublet is ninefold ($=3\times 3$) degenerate
or at least fourfold degenerate with respect to the 
first two generations. Whether such a scheme can be incorporated
in some aesthetically acceptable extension of the Standard Model or not is 
a separate question that we have not attempted to explore in this paper.

   This work was supported in part by the Director, Office of Energy Research,
Office of High Energy and Nuclear Physics, Division of High Energy Physics
of the U.S. Department of Energy under Contract No. DE-DC03-76SF00098 and 
in part by the National Science Foundation under Grant No. PHY-95-14797.

\end{document}